\documentclass[conference]{IEEEtran}
\IEEEoverridecommandlockouts

\usepackage{cite}
\usepackage{amsmath,amssymb,amsfonts}
\usepackage{algorithmic}
\usepackage{graphicx}
\usepackage{textcomp}
\usepackage{xcolor}
\usepackage{array}
\usepackage{tabularx}
\usepackage{booktabs}
\usepackage{url}

\usepackage{mdframed}
\newtheorem{mdtheorema}{Code snippet}
\definecolor{aliceblue}{rgb}{0.97, 0.98, 1.0}

\usepackage[pscoord]{eso-pic}
\usepackage{fancyvrb}
\DefineShortVerb[commandchars=\\\{\}]{\|}
\DefineVerbatimEnvironment{Highlighting}{Verbatim}{commandchars=\\\{\}}
\newenvironment{Shaded}%
  {\begin{mdframed}[backgroundcolor=aliceblue]\begin{mdtheorema}}%
  {\end{mdtheorema}\end{mdframed}}
\newcommand{\ControlFlowTok}[1]{\textcolor[rgb]{0.00,0.44,0.13}{\textbf{{#1}}}}

\newcommand{\DataTypeTok}[1]{\textcolor[rgb]{0.56,0.13,0.00}{{#1}}}
\newcommand{\DecValTok}[1]{\textcolor[rgb]{0.25,0.63,0.44}{{#1}}}

\newcommand{\CommentTok}[1]{\textcolor[rgb]{0.38,0.63,0.69}{\textit{{#1}}}}
\newcommand{\PreprocessorTok}[1]{\textcolor[rgb]{0.00,0.44,0.13}{{#1}}}

\newcommand{\ImportTok}[1]{\textcolor[rgb]{0.00,0.44,0.13}{{#1}}}

\newcommand{\NormalTok}[1]{{#1}}

\newcommand{\esp}{{\sc esp}}
\newcommand{\espml}{{\sc esp4ml}}
\newcommand{\hlsml}{{\sc hls4ml}}

\def\BibTeX{{\rm B\kern-.05em{\sc i\kern-.025em b}\kern-.08em
    T\kern-.1667em\lower.7ex\hbox{E}\kern-.125emX}}

\begin{document}

\title{ESP4ML: Platform-Based Design of Systems-on-Chip for Embedded Machine Learning 
}

\author{\IEEEauthorblockN{Davide Giri, Kuan-Lin Chiu, Giuseppe Di Guglielmo,
    Paolo Mantovani and Luca P. Carloni}
\IEEEauthorblockA{\textit{Department of Computer Science} $\cdot$
\textit{Columbia University}, New York\\
$[$davide\_giri, chiu, giuseppe, paolo, luca$]$@cs.columbia.edu}
}


\maketitle

\begin{abstract}
  We present \espml, an open-source system-level design flow to build and
  program SoC architectures for embedded applications that require the hardware
  acceleration of machine learning and signal processing algorithms.
  We realized \espml\ by combining two established open-source projects
  (\esp\ and \hlsml) into a new, fully-automated design flow.
  For the SoC integration of accelerators generated by \hlsml, we designed
  a set of new parameterized interface circuits synthesizable with high-level synthesis.
  For accelerator configuration and management, we developed an embedded software
  runtime system on top of Linux.
  With this HW/SW layer, we addressed the challenge of
  dynamically shaping the data traffic on a network-on-chip to activate
  and support the reconfigurable pipelines of accelerators that are needed by the
  application workloads currently running on the SoC.
  We demonstrate our vertically-integrated contributions with the FPGA-based
  implementations of complete SoC instances booting Linux and executing
  computer-vision applications that process images taken from the Google Street
  View database.
\end{abstract}


\section{Introduction}
\label{sec:intro}
Since 2012, when the use of deep neural networks for classifying million of
images from the web gave spectacular results~\cite{krizhevsky12,lecun15}, the
design of specialized accelerators for machine learning (ML) has become the
main trend across all types of computing systems~\cite{sze17}.
While the initial focus was mostly on systems {\em in the cloud}, the demand for
enabling machine learning into embedded devices {\em at the edge} keeps growing~\cite{deng19}.
To date, most research efforts have focused on the accelerator design in
isolation, rather than on their integration into a complete system-on-chip (SoC).
However, to realize innovative embedded systems for such domains as robotics,
autonomous driving, and personal assistance, ML accelerators must be coupled
with accelerators for other types of algorithms such as signal processing or
feedback control.
Furthermore, as the complexity of ML applications keeps growing, the challenges
of integrating many different accelerators into an SoC at design time and managing
the shared resources of the SoC at runtime become much harder.

\begin{figure}[t]
  \centering
  \includegraphics[width=0.95\columnwidth]{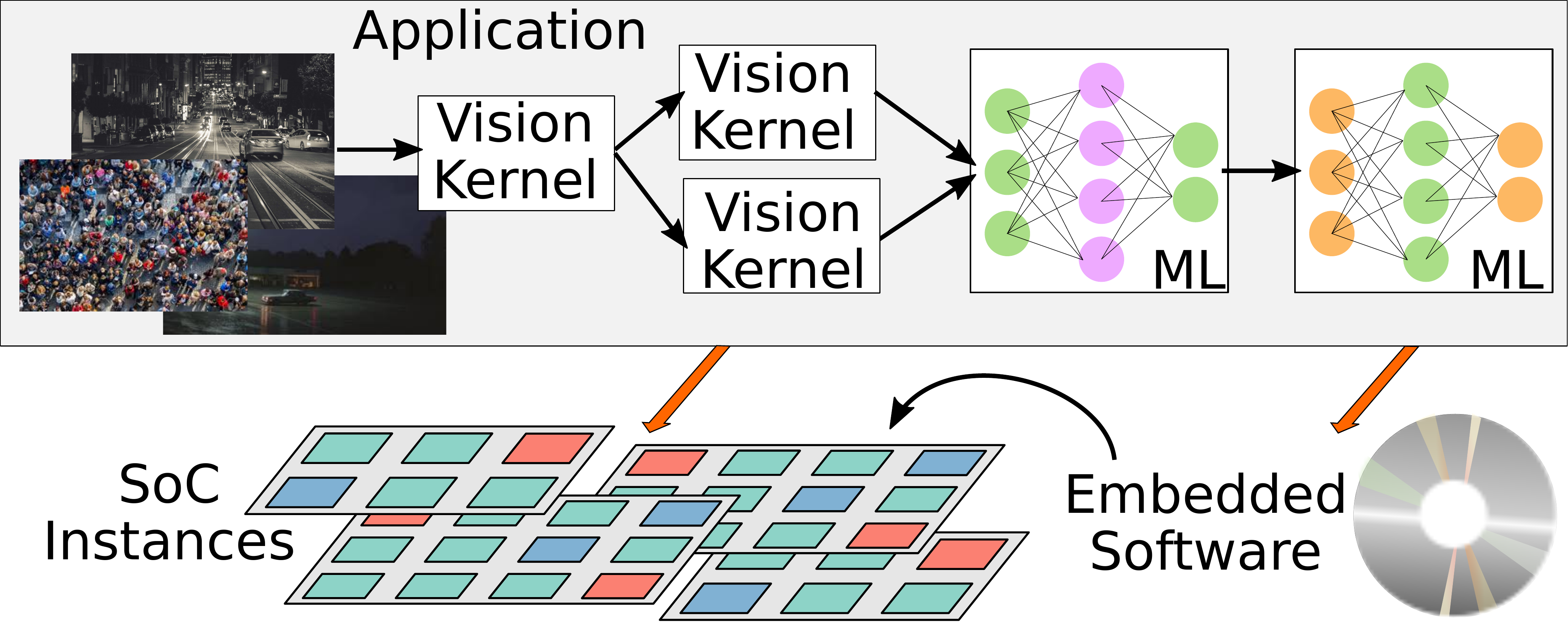}
  \caption{The proposed design flow maps full embedded applications into a
    complete SoC instance, which hosts reconfigurable pipelines of ML
    accelerators and other accelerators (e.g. for Computer Vision) connected via a NoC.}
  \vspace{-0.5cm}
  \label{fig:overview}
\end{figure}

In this paper we present \espml, a system-level design flow that enables the
rapid realization of SoC architectures for embedded machine learning.
With \espml, SoC designers can integrate at design time many heterogeneous
accelerators that can be easily connected at run-time form various
tightly-coupled pipelines (Fig.~\ref{fig:overview}).
These accelerator pipelines are reconfigured dynamically
(and transparently to the application programmer) to support the particular
embedded application that is currently running on top of Linux on the SoC processor.

To realize \espml, we embraced the concept of
{\em open-source hardware (OSH)}~\cite{gupta17}, in multiple ways.
First, our main goal is to simplify the process of designing complete SoCs that can be
rapidly prototyped on FPGA boards. The \espml\ users can focus on the design of
specific accelerators, which is simplified with high-level synthesis (HLS), while
reusing available OSH designs for the main SoC components (e.g. the Ariane
{\sc risc-v} processor core~\cite{ariane}).
Second, \espml\ is the result of combining two existing OSH projects that have
been independently developed: \esp\ and \hlsml.
\begin{itemize}
\item
\esp\ is a platform for developing heterogeneous SoCs that
    promotes the ideas of platform-based design~\cite{carloni_dac16,esp}.

\item
    \hlsml\ is a compiler that translates ML models developed with
    commonly used open-source packages such as {\sc Keras} and {\sc PyTorch}
    into accelerator specifications that can be synthesized with HLS for
    FPGAs~\cite{Duarte_2018,hls4ml}. While originally developed for research in
    particle physics, \hlsml\ has broad applicability.
\end{itemize}

To combine these two projects and reach our main goal~\footnote{We released the contributions of this paper as part of the ESP project on Github~\cite{esp}.}:

\begin{enumerate}
\item We enhanced the \esp\ architecture to support the reconfigurable activation
  of pipelines of accelerators, by implementing {\em point-to-point (p2p) communication}
  channels among them. This is done by reusing only the preexisting
  interconnection infrastructure without any overhead, i.e. without any addition of
  channel queues, routers, or links in the network-on-chip (NoC).
\item We augmented the \esp\ methodology with an {\em application programming
  interface (API)} that for a given embedded application and a target SoC
  architecture allows the specification of the software part to be
  accelerated as a simple dataflow of computational kernels.
\item We developed a {\em runtime system} on top of Linux that takes this dataflow and
  translates it into a pipeline of accelerators that are dynamically configured,
  managed, and kept synchronized as they access shared data.
  This is done in a way that is fully transparent to the application programmer.
\item We enhanced the SoC integration flow of \esp\ by designing new parameterized
  interface circuits (synthesizable with HLS) that encapsulate accelerators
  designed for Vivado HLS~\cite{vivado_hls}, without requiring any modification
  to their designs.
  This provides an adapter layer to bridge the {\em ap\_fifo} protocol from
  Vivado HLS to the \esp\ accelerator interface so that \espml\ users are only
  responsible for setting the appropriate parameters for DMA transactions (i.e.,
  transaction length and offset within the virtual address space of the
  accelerator).
\item We encapsulated \hlsml\ into a fully automated design flow that
  takes an ML application developed with {\sc Keras} TensorFlow and
  the {\em reuse factor} parameter to control parallelization
  specified within \hlsml\ and returns an accelerator that can be
  integrated within a complete SoC. This required no modification to
  the code generated with the \hlsml\ compiler.

\end{enumerate}

We demonstrate the successful vertical integration of these
contributions by presenting a set of experimental results that we
obtained with \espml.  Specifically, we designed two complete SoC
architectures, implemented them on FPGA boards, and used them to run
embedded applications, which invoke various pipelines of accelerators
for ML and computer vision.  Compared to an Intel processor, an ARM
processor, and an NVIDIA embedded GPU, energy-efficiency speedups
(measured in terms of frames/Joule) are above $100\times$ in some
cases. Furthermore, thanks to the efficient p2p-communication
mechanisms of \espml, the execution of these applications presents a
major reduction of the off-chip memory access compared to the
corresponding versions that use off-chip memory for inter-accelerator
communication, which is normally the most efficient accelerator
cache-coherence model for non-trivial workloads with regular memory
access pattern~\cite{giri_ieeemicro18}.

\section{Background}
\label{sec:back}
We give a quick overview of the \esp\ and \hlsml\ projects to provide basic
information to read the subsequent sections.

{\bf Embedded Scalable Platforms.}
\esp\ is an open-source research platform for the design of heterogeneous
SoCs~\cite{carloni_dac16}. The platform combines an architecture and
a methodology. The flexible tile-based architecture simplifies the integration
of heterogeneous components through a combination of hardware and software
sockets. The companion methodology raises the level of abstraction to
system-level design by decoupling the system integration from the design and
optimization of the various SoC components (accelerator, processors, etc.)~\cite{mantovani_aspdac16}.

The \esp\ tile-based architecture relies on a multi-plane packet-switched
network-on-chip (NoC) as the communication medium for the entire SoC.
The interface between a tile and the NoC consists of a wrapper (the hardware
part of a socket) that implements the communication mechanisms together with
other {\em platform services}.
For example, the socket of an accelerator tile typically implements:  a
configurable direct-memory access (DMA) engine, interrupt-request logic,
memory-mapped registers, and the register-configuration logic.

\begin{figure}[t]
  \centering
  \includegraphics[width=\columnwidth]{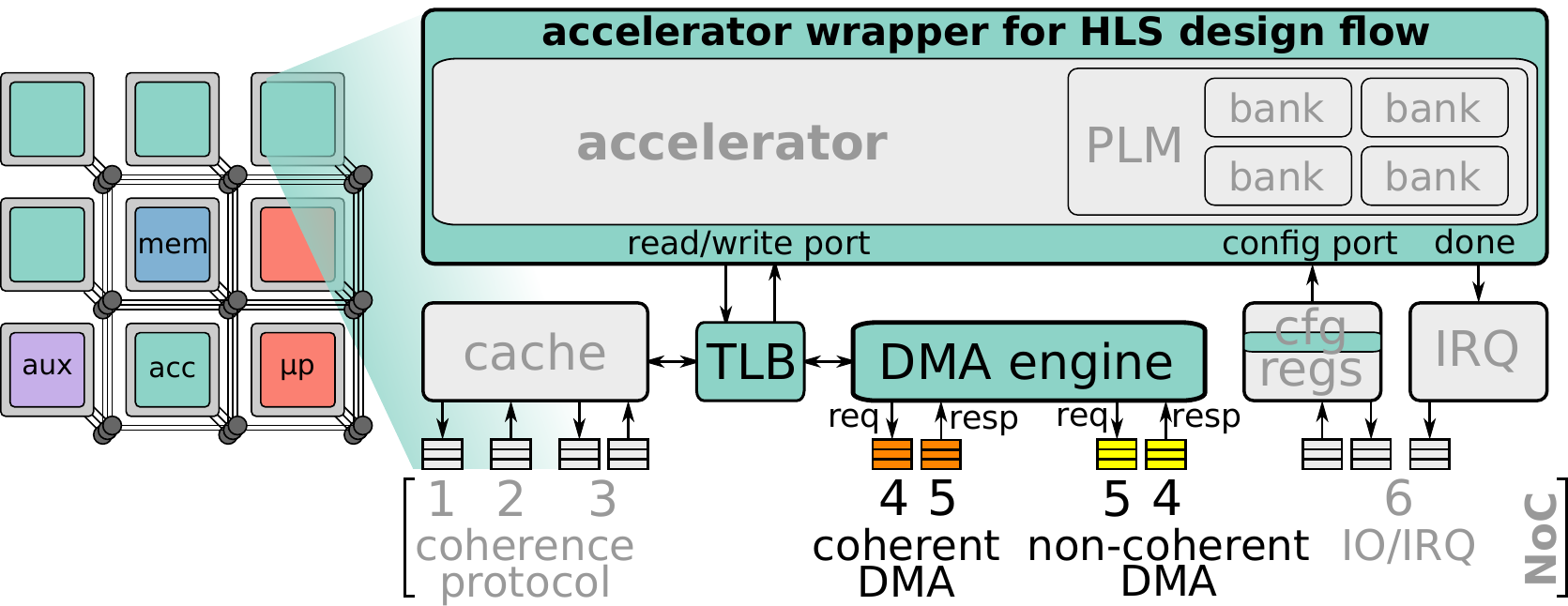}
  \caption{Example of a $ 3 \times 3 $ instance of \esp\ with zoom into the
    accelerator tile, taken from Giri et al.~\cite{giri_nocs18}.
    Components in gray with shaded text are integrated from \esp\ without
    modifications, while we modified the components in color with black text to
    support the \espml\ design flow.}
  \label{fig:esp}
\vspace{-0.3cm}
\end{figure}

In \esp, an NoC is an $ M \times N $ 2D-mesh, corresponding to a grid of tiles
of configurable size: e.g.,
Fig.~\ref{fig:esp} shows a $ 3 \times 3 $ instance of an \esp\ SoC with two
processor tiles, one memory tile, one auxiliary tile, and five accelerator tiles.
An {\em NoC plane} is a set of bi-directional links of configurable
width (e.g. 32 or 64 bits) that connect pairs of adjacent tiles in the NoC.
The \esp\ architecture allots two full planes of the NoC to the
accelerators, which use them to move efficiently long sequences of
data between their on-chip local private memories and the off-chip
main memory (DRAM). These data exchanges, called either {\em loads}
or {\em stores} depending on their direction, happen via DMA,
i.e. without involving the processor cores, which instead typically
transfer data at a finer granularity (i.e. one or few cache lines)~\cite{mantovani_cases16}.
Note that DMA requests and responses are routed through decoupled NoC planes to
prevent deadlock when multiple accelerators and multiple memory tiles are
present. 
In Section~\ref{sec:p2p} we show how we leverage this DMA queues decoupling
to efficiently implement p2p communication for \espml.

The \esp\ methodology supports a design flow that leverages SystemC and Cadence
Stratus HLS~\cite{stratus_hls} for the specification and implementation of an
accelerator to be plugged into the accelerator wrapper, as shown in Fig.~\ref{fig:esp}.
\esp\ users are responsible for the core functionality of their accelerators
and for adapting the template load/store functions provided in the synthesizable
SystemC \esp\ library.

{\bf HLS4ML.} The \hlsml\ project allows designers to specify ML models
and neural-network architectures for a specific task (e.g. image classification)
by using a common open-source software such as Keras~\cite{chollet2017keras},
PyTorch~\cite{paszke2017automatic}, and ONNX~\cite{onnx}.
A trained ML model to be used for inference is described with a couple
of standard-format files: a {\small \tt JSON} file for the
network topology and a {\small \tt HDF5} file for the model
weights and biases.  These are the inputs of the \hlsml\ compiler,
which automatically derives a hardware implementation of the
corresponding ML accelerator that can be synthesized for FPGAs using
HLS tools~\cite{nane2015survey}.  While \hlsml\ currently supports
only Vivado HLS~\cite{vivado_hls}, its approach can
be extended to other HLS tools~\cite{Duarte_2018}, possibly targeting ASIC as
well.

For an ML accelerator, the trade-offs among latency, initiation interval, and
FPGA-resource usage depend on the degree of parallelization of its inference
logic. In \hlsml, these can be balanced by setting the {\em reuse factor},
which is a single configuration parameter that specifies the number of times a
multiplier is used in the computation of a layer of neurons.

\section{The Proposed Design Flow}
\label{sec:flows}

\begin{figure}[t]
  \centering
  \includegraphics[width=0.75\columnwidth]{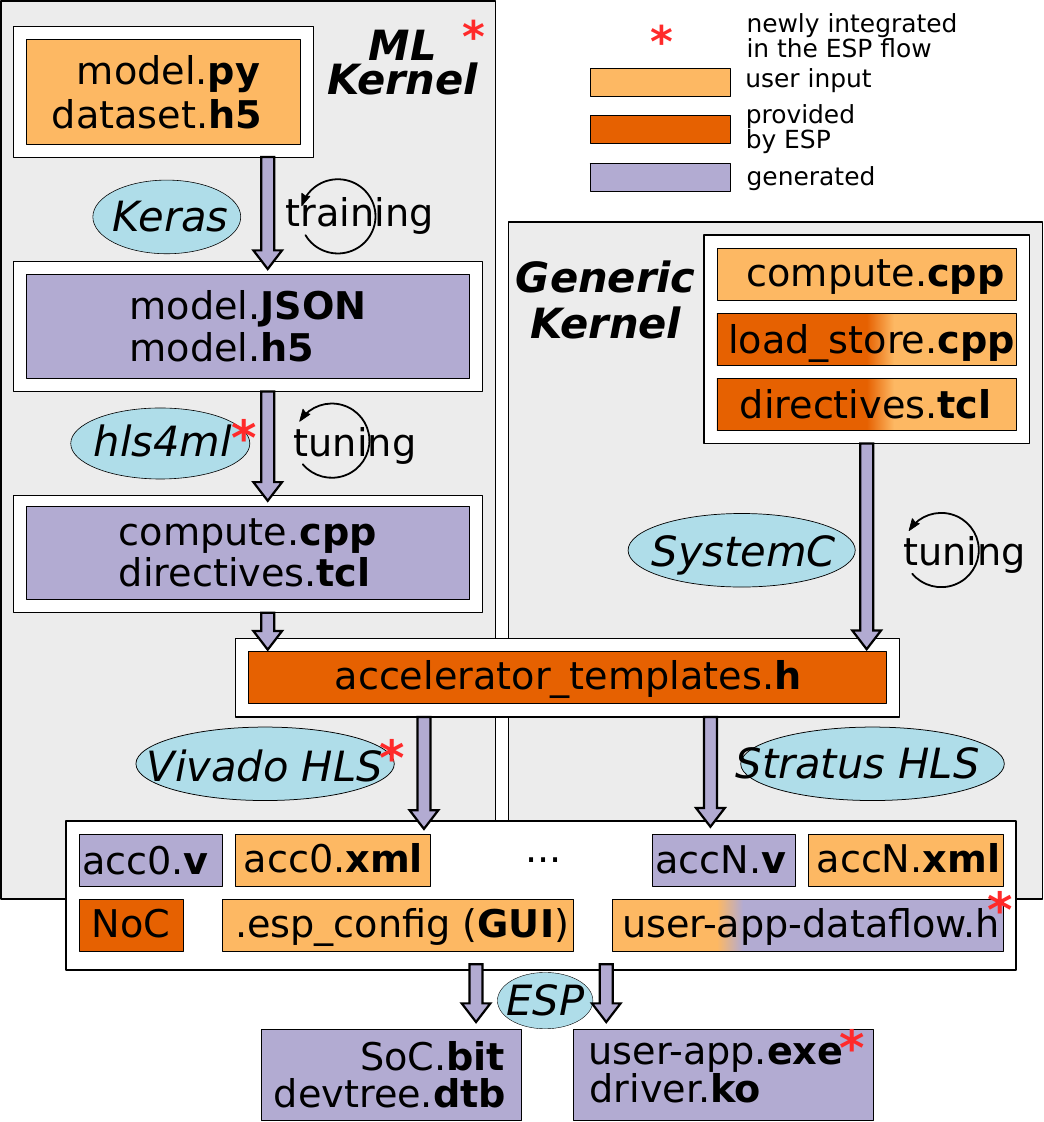}
  \caption{The proposed design flow for embedded machine learning.
  }
  \label{fig:flow}
\vspace{-0.3cm}
\end{figure}

Fig.~\ref{fig:flow} shows the \espml\ flow to design SoCs for embedded ML applications.
From the \esp\ project, we adopted the flow to design and integrate accelerators
for generic computational kernels (right) and we implemented a new flow to
design accelerators for ML applications, which leverages \hlsml\ (left).
Furthermore, we enabled the runtime reconfiguration of the communication
among accelerators through a software application (generated from a
user-specified dataflow) and a new platform service for reconfigurable p2p
communication (implemented in the wrapper of the accelerator tile).

In order to integrate accelerators compiled by \hlsml, we extended the
SoC generation flow of \esp\ to host RTL components synthesized with
Vivado HLS.  We designed a new template wrapper that is split into a
source file for Vivado HLS synthesis directives and an RTL adapter for
the \esp\ accelerator tile.  These template source files are
automatically specialized for a particular instance of ML accelerator
depending on input and output size as well as on precision and data type
(e.g. 16-bits fixed-point).

The portion of the wrapper processed by Vivado HLS implements the control logic
to make DMA transaction requests and handles the synchronization between DMA
transactions and the computational kernel. Fig.~\ref{fig:hlstop} shows the gist of
the top-level function: the {\small \tt LOAD} function gets and unpacks data from the data
read port into local memories; the {\small \tt COMPUTE} function calls the 
computational kernel (e.g. generated from \hlsml); the {\small \tt STORE}
function packs the data from local memory and pushes them to the data write
port. In addition, both {\small \tt LOAD} and {\small \tt STORE} functions set the
appropriate virtual address and length for the current transaction. This
information is computed based on the current iteration index of the main loop, the
size of the dataset and the size of the local buffers. Some of the parameters
needed are set at runtime through configuration registers (e.g.
{\small \tt conf\_size}).

\begin{figure}[t]  \scriptsize

\begin{Shaded}
\begin{Highlighting}[]
\DataTypeTok{void}\NormalTok{ TOP (word *out, word *in1, }\DataTypeTok{unsigned}\NormalTok{ conf_size,}
\NormalTok{     dma_info_t *load_ctrl, dma_info_t *store_ctrl)}
\NormalTok{\{}
\NormalTok{  word _inbuff[IN_BUF_SIZE];}
\NormalTok{  word _outbuff[OUT_BUF_SIZE];}

\NormalTok{go:}
  \ControlFlowTok{for}\NormalTok{ (}\DataTypeTok{unsigned}\NormalTok{ i = }\DecValTok{0}\NormalTok{; i < n_chunks; i++) \{}
\NormalTok{    LOAD(_inbuff, in1, i, load_ctrl, }\DecValTok{0}\NormalTok{);}
\NormalTok{    COMPUTE(_inbuff, _outbuff);}
\NormalTok{    STORE( _outbuff, out, i, store_ctrl, conf_size);}
\NormalTok{  \}}
\NormalTok{\}}
\end{Highlighting}
\end{Shaded}

\caption{Example of top-level function of the \esp~wrapper for Vivado HLS.} \vspace{-0.5cm} \label{fig:hlstop} \end{figure}

The RTL portion of the wrapper includes a set of shallow FIFO queues that
decouple the control requirements of the FIFO interface in Vivado HLS from the
protocol of the accelerator tile in \esp. In addition to FIFO queues, the wrapper
binds the \esp\ configuration registers to the corresponding signals of the
accelerator, such as {\small \tt conf\_size} in Fig.~\ref{fig:hlstop}.
The list of registers is specified into an {\small \tt XML} file for each
accelerator following the default \esp\ integration flow.

\section{Point-to-Point Communication Services}
\label{sec:p2p}

Section~\ref{sec:flows} explains how \espml\ users can specify the
accelerators for their target embedded applications. Once these are
implemented as RTL intellectual property (IP) blocks, the \esp\
graphic configuration interface can be used to pick the location of
each accelerator in the SoC and generate the appropriate hardware
wrappers, including routing tables, and Linux device drivers.  The
\esp\ infrastructure then generates a bitstream for Xilinx FPGAs and a
bootable image of Linux that can run on the embedded {\sc risc-v}
processor in the ESP SoC~\cite{mantovani_dac16}.

The \esp\ design flow, however, used to lack the ability to map the application dataflow
onto the user-level software and to dynamically reconfigure the NoC routers
to remap DMA transactions onto p2p data transfers among accelerators.
Hence, we developed a new p2p platform service for \esp\ architectures that is
compatible with the generic accelerator tile wrapper.

\begin{figure}[t]  \scriptsize

\begin{Shaded}
\begin{Highlighting}[]
\PreprocessorTok{#include }\ImportTok{"libesp.h"}
\PreprocessorTok{#include }\ImportTok{"dflow1.h"}

\DataTypeTok{int}\NormalTok{ main(}\DataTypeTok{int}\NormalTok{ argc, }\DataTypeTok{char}\NormalTok{ **argv)}
\NormalTok{\{}
    \DataTypeTok{int}\NormalTok{ errors = }\DecValTok{0}\NormalTok{;}
\NormalTok{    contig_handle_t contig;}
    \DataTypeTok{uint8_t}\NormalTok{ *buf;}

    \CommentTok{// Allocate memory}
\NormalTok{    buf = (}\DataTypeTok{uint8_t}\NormalTok{*) esp_alloc(&contig, DATASET_SIZE);}

    \CommentTok{// Initialize buffer}
\NormalTok{    init_buffer(buf);}
    
    \CommentTok{// Execute accelerator(s) dataflow.}
    \CommentTok{// The configuration specifies the communication}
    \CommentTok{// for each accelerator invocation: DMA or P2P.}
\NormalTok{    esp_run(dflow1_cfg, NACC);}

    \CommentTok{// Validation}
\NormalTok{    errors += validate_buffer(buf);}

    \CommentTok{// Free memory}
\NormalTok{    esp_cleanup();}

    \ControlFlowTok{return}\NormalTok{ errors;}
\NormalTok{\}}
\end{Highlighting}
\end{Shaded}

\vspace{-0.2cm} \caption{Generated \espml~code to spawn multiple HW-accelerated threads.} \vspace{-0.5cm} \label{fig:main} \end{figure}

First, we defined two additional registers common to all accelerators.
The {\small \tt LOCATION\_REG} is a read-only register that exposes the
x-y coordinates of an accelerator on the NoC to the operating system.
The {\small \tt P2P\_REG} is the p2p configuration register, which holds the
following information: p2p store is enabled, p2p load is enabled, number of
source tiles (1 to 4) for the load transactions, x-y coordinates of the source tiles.
We also modified the \esp\ device driver such that any registered accelerator,
(discovered when {\small \tt probe} is executed) is added to a global linked list
protected by a {\small \tt spinlock}. This list allows any thread executing the
code of an accelerator device-driver in kernel mode to access information
related to other accelerators. Since this information includes the base address
of the configuration registers, a device name, already known in user space, can
be mapped to the corresponding x-y coordinates. These coordinates are not
exposed to user space and the application dataflow can be specified by simply using
the accelerator names. Hence, the application is completely independent from the
particular SoC floorplan.

To support accelerator p2p transactions we made minor modifications to
translation-lookaside buffer (TLB) and DMA
controller in the ESP accelerator tile wrapper~\cite{mantovani_cases16}. A key aspect of our
implementation is that all p2p transactions are on-demand, that is
they must be initiated by the receiver. The sender accelerator tile
waits for a p2p load request before forwarding data to the NoC.
Implementing p2p stores on-demand is necessary to prevent long packets
of data being stalled in the NoC links while the accelerator
that is downstream in the dataflow is not ready to accept them. For the same
reason our solution guarantees the ``consumption assumption''~\cite{song03} for all
supported dataflow configurations. An accelerator tile will only
request data when it has enough space to store it locally.

This mechanism is completely transparent to the
accelerator, which still operates as if regular DMA transactions were
to occur, while performance and energy consumption largely benefit
from close-distance communication and a drastic reduction in accesses
to DRAM or to the last-level cache.

We built this p2p communication service without adding any NoC planes, nor
queues at the NoC interface, because we rely on queues that are otherwise unused
for regular DMA transactions.
Specifically, we carefully reused available queues in the \esp\
accelerator tile.

\section{Runtime System for Accelerators}

After implementing the p2p service, we developed a software API to hide the
details of memory allocation, accelerator invocation, and synchronization from
user-space software. Dependencies across accelerators are specified through a
simple dataflow.
By modifying a template that is automatically generated for the given
SoC architecture, the \espml\ users can define a dataflow of
accelerator invocations. For each invocation they can specify whether
to use DMA or p2p communication and they can set other
accelerator-specific communication parameters.

The snippet in Fig.~\ref{fig:main} shows an example of automatically generated
applications that reads two dataflow configurations from {\small \tt dflow1.h}.
For each configuration the application spawns as many threads as the number of
running accelerators to exploit all the available parallelism in the dataflow.
Since accelerators that use the p2p service are automatically synchronized in
hardware, the software runtime incurs minimal overhead. This is limited to the
{\small \tt ioctl} system calls that are used to start the accelerators~\cite{mantovani_cases16}.
When \espml\ users set the dataflow parameters to use DMA only, dependencies are
enforced with {\small \tt pthread} primitives.
Thanks to our software runtime, \espml\ users can dynamically reshape the data
traffic on the NoC to activate a reconfigurable pipeline of accelerators for the
given embedded application
In addition, they can tune the throughput of the system by balancing each stage
of this pipeline: e.g., if a slow accelerator is feeding a faster one, multiple
instances of the slower accelerator can be activated to feed a single
accelerator downstream.

\section{Experimental Results}
\label{sec:eval}

\begin{figure}
  \centering
  \includegraphics[width=1\columnwidth]{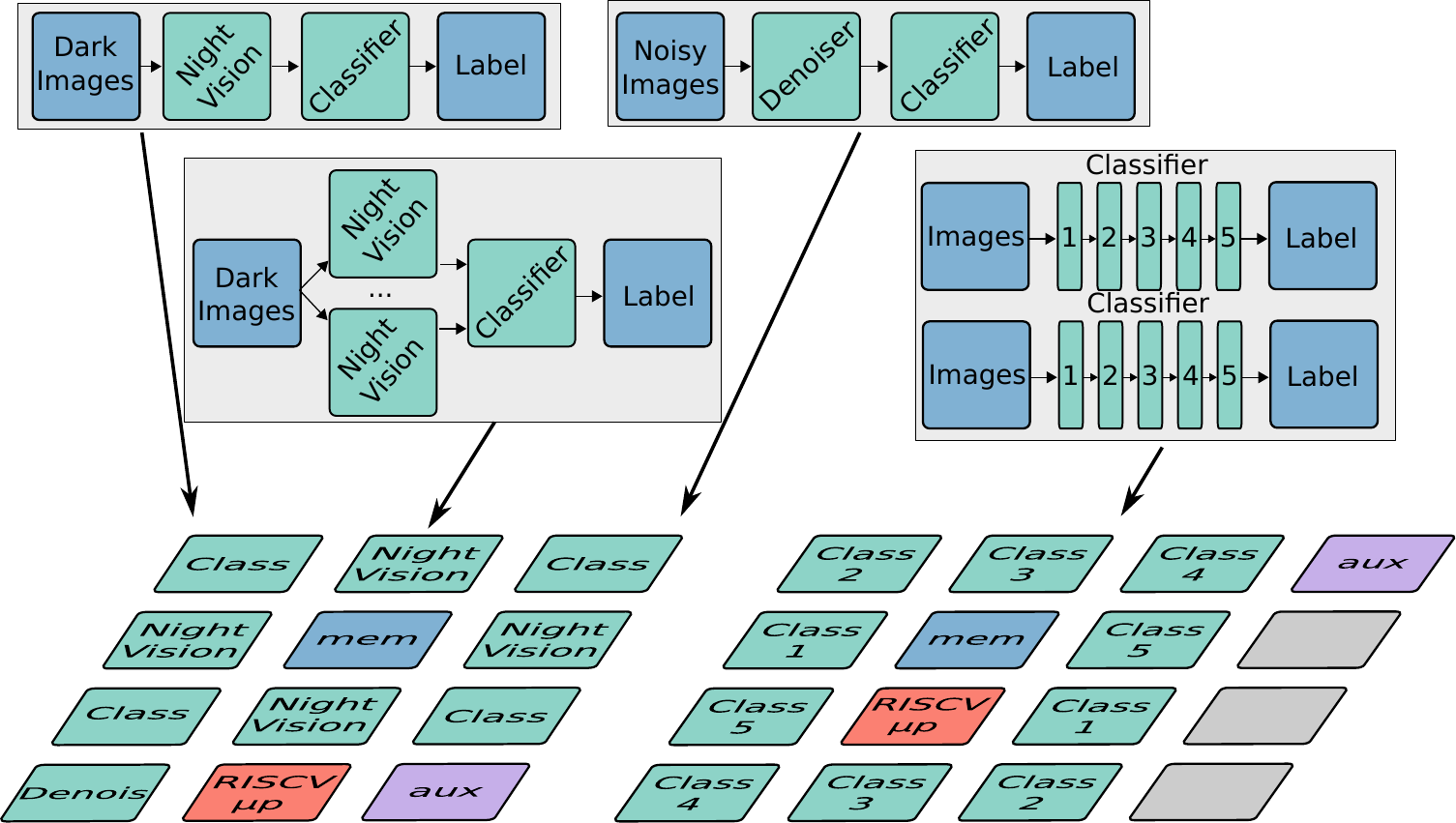}
\vspace{-0.1cm}
  \caption{Dataflow of the four case-study applications discussed in the
    evaluation and corresponding mapping onto two instances of ESP.}
  \label{fig:accelerators}
\vspace{-0.4cm}
\end{figure}

{\bf Applications.}
Street View House Numbers (SVHN) is a real-world image dataset obtained from
Google Street View pictures~\cite{svhnDataset}.
SVHN is similar to the MNIST dataset, but it is ten times bigger (600,000 images split
in training, test, extra-training datasets). For SVHN, the problems get
significantly more laborious due to the environmental noise in the pictures
(including shadows and distortions).
We developed two embedded applications for the SVHN dataset: digit classification and
image denoising. For both, we adopted ML solutions and trained our models in
{\sc Keras}.
Recalling the \espml\ flow overview of \figurename~\ref{fig:overview}, the upper part
of \figurename~\ref{fig:accelerators} shows concrete instances for these two applications.

For the digit classification problem, we defined a Multilayer Perceptron (MLP)
with four hidden layers.
The size of the fully connected network is 1024x256x128x64x32x10.
We used dropout layers with a 0.2 rate to prevent overfitting during
training. The trained model accuracy is 92\%.
For the denoising problem, we designed an autoencoder model.
The network size is 1024x256x128x1024, and the compression factor in the
bottleneck is 8.
We added Gaussian noise to the SVHN dataset and trained the model with a 3.1\%
reconstruction error.

We also developed one application outside the ML domain, which is a night
computer vision application consisting of three kernels: noise filtering,
histogram, and histogram equalization.
For the purpose of this evaluation, we darkened the SVHN dataset and we used
this Night-Vision application as a pre-processing step of the MLP classifier
described above.

\begin{table}[t]
\scriptsize
\centering
\caption{Summary of results using the best-case configuration}\label{tab:transposed_results}
\vspace{-0.2cm}
\begin{tabular}{l | r | r | r }
\toprule
   & {\bf \textsc{NightVision \&}} & {\bf \textsc{Denoiser \&}} & {\bf \textsc{Multi-tile}} \\
   & {\bf \textsc{Classifier}} & {\bf \textsc{Classifier}} & {\bf \textsc{Classifier.}} \\
\midrule
  {\sc LUTs}             & 48\%   & 48\% & 19\%     \\
  {\sc FFs}              & 24\%   & 24\% & 11\%     \\
  {\sc BRAMs}            & 57\%   & 57\% & 21\%     \\
  {\sc Power (W)}        & 1.70   & 1.70 & 0.98     \\
  {\sc Frames/s \espml}  & 35,572 & 5,220 & 28,376  \\
  {\sc Frames/s Intel i7}& 1,858  & 30,435 & 82,476 \\
  {\sc Frames/s Jetson}  & 377    & 2,798 & 6,750   \\
  \bottomrule
\end{tabular}
\vspace{-0.5cm}
\end{table}

{\bf Accelerators and SoCs.} We designed two SoCs that we synthesized for FPGA
with the \espml\ flow. As shown in \figurename~\ref{fig:accelerators},
these SoCs contain many (up to ten) accelerators for the target applications
and one Ariane {\sc risc-v} core. Table~\ref{tab:transposed_results} shows the FPGA
resources usage and the dynamic power dissipation as reported by Xilinx Vivado.
We designed the Classifier and the Denoiser with {\sc Keras} and we compiled them
with \hlsml\ within the \espml\ flow. We then designed a partitioned
version of the Classifier, by distributing the computation across five
accelerators. Finally, we designed the accelerator for the Night-Vision
kernels by leveraging another HLS-based design flow within \esp:
i.e., we designed them in SystemC and synthesized them with \emph{Cadence Stratus HLS}.

{\bf Experimental Setup.}
We implemented the two \espml\ SoCs of
\figurename~\ref{fig:accelerators} on a Xilinx Ultrascale+ FPGA board
with a clock frequency of 78MHz. We ran all the experiments by using
this board and executing the test embedded applications on top of
Linux running on the Ariane core.
We compared the execution of these applications on the \espml\ SoC with the
hardware accelerators versus the execution of the same applications in software
on the following two platforms:
(a) an Intel i7 8700K processor and (b) an NVIDIA Jetson TX1 model, which is an
embedded system that combines a 256-core NVIDIA Maxwell GPU with a Quad-Core
ARM Cortex-A57 MPCore.
Based on the available datasheet, we considered values of power consumption
equal to $1.5W$ and $10W$ for the ARM core and the GPU, respectively. For the
Intel core, we estimated a TDP of $78.6W$ (the nominal value is $95W$).

{\bf Results.}
The three bottom lines of Table~\ref{tab:transposed_results} report the
performance of the three platforms measured in terms of processed frames per
second. The FPGA implementations of the SoC designed with \espml\ offer better
performance compared to a commercial embedded platform like the Jetson TX1.
The Intel i7 cores predictably provides the best performance, aside for the
case of the Night-Vision application, which is a single-threaded program.

\begin{figure}[t]
  \centering
  \includegraphics[width=1\columnwidth]{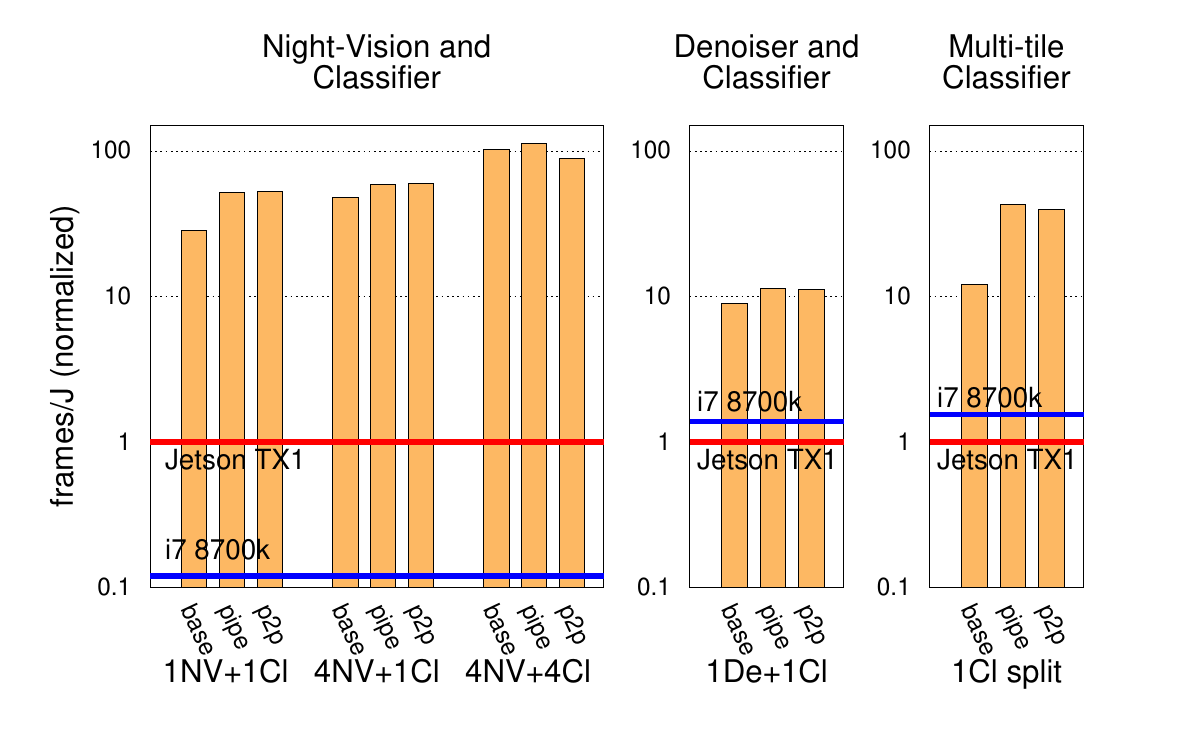}
   \vspace{-0.8cm}
  \caption{Energy efficiency in terms of $frames/Joules$ for the
    \espml\ compared to an NVIDIA embedded GPU and an Intel i7 core.
    The three bars show the performance of two \espml\
    techniques: reconfigurable pipelines of accelerators w/
    (\emph{p2p}) and w/o (\emph{p2p}) p2p communication.
  }
  \label{fig:energy}
\vspace{-0.4cm}
\end{figure}

\figurename~\ref{fig:energy} compares the execution of the applications on the
three platforms in terms of energy efficiency, measured as $frames/Joule$ (in
logarithmic scale).
Notice that all the accelerator execution-time measurements include the overhead
of the \espml\ runtime system managing the accelerators invocations as well as
the overhead of the accelerators Linux device drivers.
The horizontal blue and red lines show the efficiency of the CPU and GPU,
respectively. For the purpose of this comparison, we report the average dynamic
power consumption for the two \espml\ SoCs as estimated by Xilinx Vivado for the
whole SoC (i.e. not just for the accelerators active in a specific test).
This is a conservative assumption, particularly if one considers that the power
consumption depends on the choice of the FPGA and that a Xilinx Ultrascale+ is a
particularly large FPGA.
Still, the \espml\ SoCs outperforms both the GPU and the CPU across all three
applications, yielding in some cases an energy-efficiency gain of over $100\times$.

\begin{figure}[t]
  \centering
  \includegraphics[width=0.85\columnwidth]{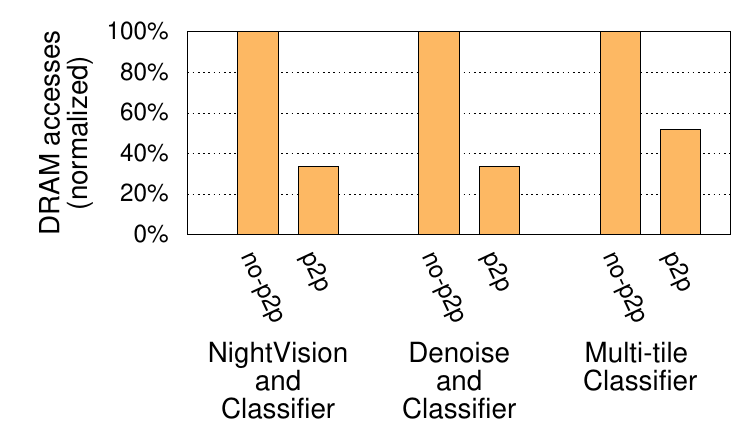}
\vspace{-0.1cm}
  \caption{Relative number of DRAM accesses w/ and w/o point-to-point
    communication among accelerators for three test applications.
    The energy savings due to a reduced access to memory are the main
    benefit of the point-to-point communication among accelerators.
\vspace{-0.5cm}
  }
  \label{fig:dram}
\end{figure}

Each cluster of bars in \figurename~\ref{fig:energy} represents an execution
based on a different pipeline of accelerators, with the number of accelerators
varying from two to eight.
The left bar of each cluster shows results for the case where the accelerators
are invoked serially in a single-thread application.
The middle bars (label \emph{pipe}) correspond to concurrent executions in a
reconfigurable pipeline, as the accelerators are invoked with a multi-threaded
application (one thread per accelerator). The right bar adds the \espml\ p2p
communication to this pipeline execution.
The results for the Night-Vision and Classifier show that the performance
increases significantly when the accelerators work concurrently in pipeline.
While p2p communication does not provide a major gain in performance in this case, its main
benefit is the reduction of off-chip memory accesses, which translates into a
major energy saving: as shown in \figurename~\ref{fig:dram}, this reduction
varies between $2\times$ and $3\times$ for the target applications.

\section{Related Work}
\label{sec:related}
As efforts in accelerators for ML continue to grow, HLS is recognized as a
critical technology to build efficient optimization flows~\cite{zhang_2017}.
For instance, Hao et al. recently proposed a PYNQ-ZI based approach to design deep
neural network accelerators~\cite{hao_2019}.
Meanwhile, various optimization techniques to deploy deep neural networks on
FPGA have been proposed~\cite{wang_2016,zhang_iccad16,zhang_2017,hao_2018}.
In this context, \hlsml~\cite{Duarte_2018} is being increasingly adopted by research
organizations and is raising interest in the industry~\cite{xilinx_cern,fastmachinelearning}.
To date, however, most open-source projects focus on the design of accelerators
in isolation. Instead, we propose the first automated open-source design flow
that leverages \esp\ and \hlsml\ to integrate multi-accelerator pipelines into SoCs.
The ESP project initially focused on the integration of generic
accelerators specified in SystemC that could operate in pipeline
through shared memory~\cite{mantovani_aspdac16}. The \espml\ flow augments ESP with the
support of accelerators designed also with common ML API and enable
runtime reconfiguration of pipelines with efficient p2p communication.

\section{Conclusions}
\label{sec:conclusions}
\espml\ is a complete system-level design flow to implement SoCs for
embedded applications that leverage tightly-coupled pipelines of many
heterogeneous accelerators.
We realized \espml\ by building on the prior efforts of two distinct
open-source projects: \esp\ and \hlsml.
In particular, we augmented \esp\ with a HW/SW layer that enables the
reconfigurable activation of accelerators pipelines through efficient
point-to-point communication mechanisms.
In addition, we built a library of
interface circuits that allow for the first time to integrate \hlsml\
accelerators for machine learning into a complete SoC using only open-source
hardware components.
We demonstrated our work with the FPGA implementations of various SoC instances
running computer-vision applications.
{\em   }

\vspace{0.04in}

{\footnotesize
  {\bf Acknowledgments}.
  This work was supported in part by DARPA (C\#: FA8650-18-2-7862) and
  in part by the National Science Foundation (A\#: 1764000). The views
  and conclusions contained herein are those of the authors and should
  not be interpreted as necessarily representing the official policies
  or endorsements, either expressed or implied,of Air Force Research
  Laboratory (AFRL) and Defense
  Advanced Research Projects Agency (DARPA) or the U.S. Government.\\
  We thank the developer team of {\em hls4ml}. We acknowledge the Fast
  Machine Learning collective as an open community of multi-domain
  experts and collaborators. This community was important for the
  development of this project.
}
 
{\footnotesize
\bibliographystyle{IEEEtran}
\bibliography{IEEEabrv,paper}
}

\end{document}